\documentclass[journal]{IEEEtran}

\ifCLASSINFOpdf
\else
\fi
 \usepackage[caption=false,font=footnotesize]{subfig}
\usepackage{cite}
\usepackage{amsmath,amssymb,amsfonts}
\usepackage{algorithmic}
\usepackage{graphicx}
\usepackage{textcomp}
\usepackage{xcolor}
\usepackage{enumitem}
\usepackage{nameref}
\usepackage{url}
\usepackage{todonotes}
\usepackage{placeins}
\usepackage{tabularx}
\usepackage{booktabs}
\usepackage{xurl}
\usepackage{booktabs}
\usepackage{makecell}
\usepackage{siunitx}
\newcommand{\ForAll}[1]{#1}
\def\BibTeX{{\rm B\kern-.05em{\sc i\kern-.025em b}\kern-.08em
    T\kern-.1667em\lower.7ex\hbox{E}\kern-.125emX}}

\begin{document}
%
%
%
%

\title{A Dynamic Capacity Allocation Model for DERs under Non-Firm Connection Agreements}

\author{Neda~Vahabzad,~\IEEEmembership{Student Member, IEEE},
        Kenneth~Bruninx,~\IEEEmembership{Member, IEEE},
        Peter~Palensky,~\IEEEmembership{Senior Member, IEEE},
        and~Pedro~P.~Vergara,~\IEEEmembership{Senior Member, IEEE}%
\thanks{N. Vahabzad, P. P. Vergara, and P. Palensky are with the Department of Electrical Sustainable Energy, Delft University of Technology, Delft, The Netherlands (e-mail: \{n.vahabzad, p.p.vergarabarrios, p.palensky\}@tudelft.nl).}%
\thanks{K. Bruninx is with the Faculty of Technology, Policy and Management, Section Energy \& Industry, Delft University of Technology, Delft, The Netherlands (e-mail: k.bruninx@tudelft.nl).}%
\thanks{This research is funded by the EU Horizon 2020 MAGPIE Project No.~101036594.}%
}
%
%

\markboth{}%
{Shell \MakeLowercase{\textit{et al.}}: Bare Demo of IEEEtran.cls for IEEE Journals}
%



\maketitle

\begin{abstract}
The growing penetration of distributed energy resources (DERs) intensifies congestion in distribution networks by introducing bidirectional power flows and increasing competition for limited network capacity, underscoring the need for effective and efficient congestion management, including flexible grid-access schemes. This paper proposes a bilevel optimization model for the dynamic allocation of connection capacity to DERs under non-firm connection agreements, aligning the objectives of distribution system operator (DSO) and DER owners. The upper-level problem, representing the DSO, determines the allocated connection capacity for all DERs, defined as maximum time-varying power limits, subject to distribution system constraints and the last-in-first-out (LIFO) allocation rule. The lower-level problem, representing DER owners, maximizes the profit of each DER within the allocated power limits. The proposed model is tested on a modified CIGRE medium-voltage (MV) network, demonstrating a balanced trade-off between grid utilization and economic efficiency. Furthermore, the model enhances DER integration, enforces transparent allocation rules, reduces variability in allocation patterns, and achieves up to an 80\% reduction in total curtailment costs compared with benchmark methods.
\end{abstract}
\noindent\textbf{\large Nomenclature}

\vspace{1mm}
\begin{tabular}{@{}p{2cm}l}
\textbf{Sets /\ Indices} \\
$\mathcal{T}$ & Set of time steps, indexed by $t$ \\
$\mathcal{N}$ & Set of buses, indexed by $i,j,k$ \\
$\mathcal{L}$ & Set of lines, indexed by $ij$ \\[1ex]

\textbf{Parameters} & \\
$\lambda_t$ &  Day-ahead electricity price (€/MWh) \\
$c^{\text{CU}}$ & Curtailment penalty coefficient (€/MWh) \\
$P_{i,t}^{\text{C}}$ & DERs connected capacity (MW) \\
$\bar{P}_{t}^{\text{G}}$ & Maximum grid capacity at time $t$ (MW) \\
$\bar{I}_{ij}$ & Thermal current limit of line $(i,j)$ (kA) \\
$\gamma$ & Line overloading penalty coefficient (€/h)\\

\IEEEpubidadjcol 

$L_{\text{S}}$ & Safe line loading ratio (p.u.) \\
$P_{i,t}^{\text{L}}$ & Active power demand (MW)\\
$Q_{i,t}^{\text{L}}$ & Reactive power demand (MVAr) \\
$R_{ij}$ & Line resistance ($\Omega$) \\
$X_{ij}$ & Line reactance ($\Omega$) \\
$\underline{V}$ & Minimum bus voltage magnitude (p.u.) \\
$\overline{V}$ & Maximum bus voltage magnitude (p.u.) 
\end{tabular}

\begin{tabular}{@{}ll}
\textbf{Variables} & \\
$P_{i,t}^{\text{A}}$ & Allocated power limit to DERs (MW) \\
$C^{\text{G}}(t)$ & Grid usage penalty cost (€/h) \\
$P_{i,t}^{\text{IN}}$ & Actual power injection from DERs (MW) \\
$L_{ij,t}$ & Line loading ratio (p.u.) \\
$I_{ij,t}$ & Line current magnitude (kA) \\
$P_{ij,t}$ & Active power flow in lines (MW) \\
$P_{ij,t}^{\text{LS}}$ & Active power losses in lines (MW) \\
$P_{i,t}^{S}$ & Slack bus active power injection (MW) \\
$Q_{ij,t}$ & Reactive power flow in lines (MVAr) \\
$Q_{ij,t}^{\text{LS}}$ & Reactive power losses in lines (MVAr) \\
$Q_{i,t}^{S}$ & Slack bus reactive power injection (MVAr) \\
$V_{i,t}$ & Bus voltage magnitude (p.u.) 
\end{tabular}

%
\IEEEpeerreviewmaketitle

\section{Introduction}
%
%
%
%
\IEEEPARstart{M}{any} distribution networks currently face congestion, either when current levels approach or exceed the thermal limits of lines and transformers, or when voltage magnitudes deviate from acceptable ranges. The increasing penetration of distributed energy resources (DERs) further intensifies these issues by introducing reverse power flows, increased line loading, and voltage fluctuations \cite{ref1}. Such conditions reduce system efficiency and increase operational costs, underscoring the need for effective and efficient congestion management strategies to ensure secure and economically efficient grid operation \cite{ref2}.

Beyond network upgrades and reconfiguration, leveraging DER flexibility to utilize existing network capacity more efficiently is increasingly being considered as a congestion management approach \cite{refnew3}. DER flexibility is activated primarily through market-based methods, which rely on economic signals or contractual mechanisms, such as dynamic tariffs, distribution capacity markets, and flexibility services, to incentivize the desired behavior from network users \cite{refnew4,ref2}. Several studies have proposed market architectures that enable interactions between DSOs and DERs for trading flexibility services and enabling efficient congestion management \cite{refnew6}. However, in practice, participation from DER owners is often limited due to insufficient economic incentives or unfavorable market conditions, resulting in low market liquidity and underutilization of available flexibility \cite{refnew11}. Moreover, reliance on market outcomes introduces uncertainty regarding the availability of flexibility, which can compromise the predictability required for secure network operation and complicate real-time congestion management by DSOs \cite{refnew12}.


Another promising congestion management approach involves non-firm or flexible connection agreements, which are contractual arrangements that provide DERs and flexible loads with conditional access to network capacity based on system conditions, typically under predefined capacity limits or curtailment rules. One of the key challenges in implementing such agreements is allocating the available network capacity among multiple holders of non-firm connection agreements under network stress conditions, as different allocation strategies favor different stakeholders and affect the overall attractiveness of these contracts \cite{ref8}. The capacity allocation problem is inherently multi-dimensional, involving trade-offs between transparency, economic efficiency, and ease of implementation \cite{ref5}. Since non-firm connection agreements constitute the main focus of this paper, these agreements and the associated allocation challenges are further discussed below.

\subsection{Non-firm Connection Agreements}
Non-firm connection agreements provide grid access based on real-time or forecasted network conditions. Such connections are typically supported by continuous monitoring of network states and dynamic adjustment of DER outputs to ensure system reliability. By enabling conditional access to available network capacity, non-firm connections can significantly reduce connection costs and lead times while increasing the overall hosting capacity \cite{refnew10}. 
However, most customers prefer firm connections due to the uncertainty and risks associated with non-firm contracts, unless substantial tariff reductions are offered \cite{ref7}. The effectiveness of non-firm agreements relies on maintaining a balance: insufficient incentives limit customer participation, while excessive ones risk market gaming and unfair cost distribution, particularly affecting firm customers. Furthermore, since capacity is shared among participants, increasing the number of participants reduces each customer’s share, raising concerns regarding the underlying capacity allocation strategy. Due to their operational complexity and reliance on user flexibility, non-firm grid connection schemes are currently mainly applied to commercial and industrial customers \cite{ref18}.

In this context, allocating the available network capacity among participants becomes a central challenge. Existing allocation methods can broadly be categorized into two groups: \textit{optimization-based} and \textit{rule-based prioritization} approaches. Optimization-based approaches provide a grid-feasible benchmark solution by maximizing allocable capacity within network limits. However, they do not explicitly define transparent allocation rules \cite{papernanda}. In contrast, rule-based prioritization methods emphasize transparent and predictable allocation rules from the perspective of DER owners and incorporate fairness principles in different ways \cite{ref18}.  
However, this often comes at the cost of reduced allocable capacity compared to optimization-based approaches, as fairness-driven prioritization rules may limit the efficient utilization of network capacity. Poorly designed prioritization rules can further exacerbate this effect, leading to additional reductions in hosting capacity and the amount of power delivered to the grid \cite{ref9}. As highlighted in \cite{ref10}, rules such as last-in first-out (LIFO), where DERs with earlier connection agreements are prioritized over those with newer agreements, can significantly affect both network operation and perceived fairness. This underscores the need to understand how different priority-based schemes influence allocation outcomes under congested grid conditions. 

The effectiveness of non-firm capacity allocation models can be evaluated from different perspectives. From the DSO perspective, non-firm grid contracts should enable higher utilization of existing infrastructure and maximize net exports at interconnection points \cite{ref14}. However, effective implementation requires allocation models that explicitly account for operational constraints, including voltage variability \cite{ref15}. In practice, non-firm agreements are most beneficial when network constraints are relatively infrequent and mild, allowing DSOs to defer costly reinforcements while limiting curtailment levels \cite{ref16}. Accordingly, DSOs assess performance based on factors such as grid utilization, curtailment levels, and voltage violations, as well as broader objectives such as loss minimization and flexibility management \cite{ref11,ref14}. From the DER (customer) perspective, non-firm connection agreements are primarily attractive due to lower tariffs \cite{ref7}; however, their viability also depends on allocation fairness, transparency, predictability, and economic efficiency, as these factors directly influence investment confidence \cite{ref17,ref9}. 



\subsection{Research Gap \& Contribution}
The development of allocation models that simultaneously address both network-level and DER-specific performance metrics under non-firm connection agreements remains an active area of research. Existing approaches typically maximize the utilization of total allocable capacity within network operational limits without explicitly considering DER-level aspects, such as economic value, transparency of the allocation mechanism, and stability of allocated capacity profiles, or rely on predefined prioritization rules that, although transparent, do not adequately account for their impact on overall network feasibility and economic efficiency.

To address this gap, this paper proposes a bilevel optimization model for dynamic allocation of connection capacity to DERs under non-firm connection agreements. The model determines time-varying upper bounds on DER power injections to ensure network feasibility, alleviate congestion, and achieve temporally stable and economically efficient allocation profiles. By explicitly representing the objectives of the DSO (upper level) and DER owners (lower level), the model captures the interaction between technical network constraints and economic incentives, thereby aligning operational decisions across stakeholders. 


The main contributions of this paper are as follows:
\begin{itemize}
\item A bilevel optimization model is developed for the dynamic allocation of grid connection capacity to DERs, integrating OPF-based network feasibility with DER-level economic decision-making. For the first time, the LIFO priority rule is incorporated directly as an optimization constraint, allowing the maximum utilization of available grid capacity while ensuring transparency of the allocation rule within a unified model. The proposed model is formulated as a time-coupled optimization framework over a one-day scheduling horizon with hourly resolution. This formulation captures the evolution of network constraints and DER behavior while limiting abrupt variations in allocated capacity. In addition, a dynamic grid-usage penalty signal defined by the DSO is incorporated to discourage network operation close to technical limits and support secure utilization of grid capacity.

\item The applicability of the proposed model to practical non-firm connection schemes is validated using a modified CIGRE medium-voltage (MV) case study with high DER penetration. The model achieves a balanced trade-off between grid feasibility, priority-based transparency, and DER economic response, while promoting temporally stable allocation profiles and mitigating excessive curtailment of lower-priority DERs. This supports the development of more practical, transparent, and economically viable non-firm connection schemes.

\end{itemize}

 

\section{Methodology}


In this section, the bilevel capacity allocation model is first introduced, with the formulation of the upper-level problem from the DSO perspective and the lower-level problem representing DER operational decisions. The solution approach adopted to solve the resulting bilevel optimization problem is then described, including the reformulation and computational aspects. Subsequently, the key performance indicators (KPIs) used to evaluate the proposed model in terms of network operation, DER integration, and economic performance are defined. Finally, two benchmark models, namely the LIFO-only and OPF-only allocation models, are introduced for comparative assessment of the proposed bilevel model.

\FloatBarrier
\subsection{Capacity Allocation Model}
\begin{figure}[t]
\centerline{\includegraphics[width=0.9\linewidth]{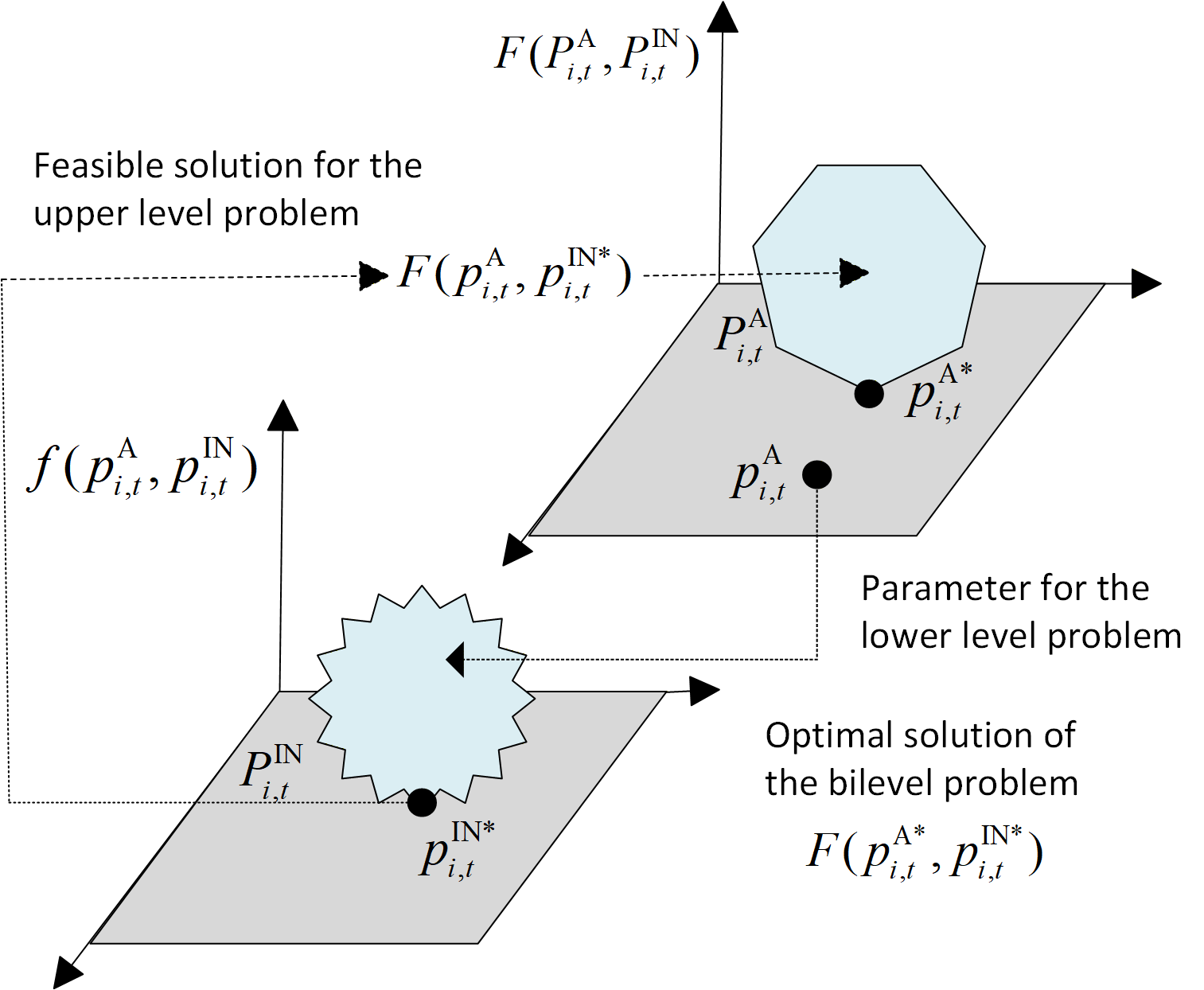}}
\caption{Schematic representation of the interaction between the upper-level and lower-level problems in the bilevel capacity allocation model.}
\label{fig:1}
\end{figure}

Figure~\ref{fig:1} provides a schematic representation of the proposed bilevel capacity allocation model, illustrating the interaction between the two decision-making levels and the coupling of their decision spaces. Since the proposed framework is formulated with hourly time resolution, the allocated connection capacity is represented as time-varying power limits corresponding to the maximum allowable DER injection during each scheduling interval. The upper level determines the power allocation limits $P^{A}_{i,t}$ for each DER and time step, which are then passed to the lower-level problem as parameters. Given these limits, each DER owner solves an individual optimization problem and determines the corresponding injected power $P^{IN}_{i,t}$ that maximizes its economic benefit. The resulting optimal injection, $P^{IN*}_{i,t}$, is then used in the upper level problem to determine whether the operating point satisfies the network constraints. Accordingly, the allocation limits $P^{A}_{i,t}$ are iteratively updated until the optimal solution $(P^{A*}_{i,t}, P^{IN*}_{i,t})$ is obtained. 
The DSO’s allocation decisions thus shape the feasible region of the DER owners’ decisions, while the DERs’ optimal responses, in turn, influence the DSO’s search for an optimal allocation limit. The bilevel capacity allocation model is formulated in the following with a one-hour time resolution.


\subsubsection{Upper-Level Problem (DSO Perspective)}
The upper-level problem (Eqs.~\eqref{eq:1}–\eqref{eq:12}) represents the DSO’s objective of supporting DER integration while mitigating congestion by minimizing both curtailment and grid-usage penalty costs, subject to network constraints, allocation rules, and operational limits:
\allowdisplaybreaks
\begingroup


\begin{IEEEeqnarray}{l}
\hspace{-1.2em}\min_{\{P_{i,t}^{\text{A}}\}}
\;\sum_{t\in\mathcal{T}} \Bigg[
C^{\text{G}}(t)
+ c^{\text{CU}}\sum_{i}(P_{i,t}^{\text{C}}-P_{i,t}^{\text{A}})
\Bigg]\Delta t
\IEEEyesnumber \label{eq:1}
\end{IEEEeqnarray}
\begin{IEEEeqnarray}{r rCl l}
\IEEEeqnarraymulticol{5}{l}{\text{s.t.}}\nonumber\\
&\sum_{i\in\mathcal{N}} P_{i,t}^{\text{A}}
&\leq&
\bar{P}_t^{\text{G}}
& \ForAll{\forall\, t\in\mathcal{T}}
\IEEEyesnumber \label{eq:2} \\[-2pt]
&
P_{i,t}^{\text{A}}
&\leq&
\min_{k\in\mathcal{H}(i)}
\left(
P_{i,t}^{\text{C}} \cdot \frac{P_{k,t}^{\text{A}}}{P_{k,t}^{\text{C}}}
\right)
& \ForAll{\forall\, i\in\mathcal{N},\, t\in\mathcal{T}}\quad
\IEEEyesnumber \label{eq:3} \\[-2pt]
&
C^{\text{G}}(t)
&=&
c_0 + \sum_{(ij)\in\mathcal{L}} f\!\left(L_{ij,t}\right)
& \ForAll{\forall\, t\in\mathcal{T}}
\IEEEyesnumber \label{eq:4} \\[-2pt]
&
P_{i,t}^{\text{L}}
&=&
P_{i,t}^{\text{IN}} 
+ \sum_{ki\in\mathcal{L}} P_{ki,t}\nonumber
&
\\
& & &
- \sum_{ij\in\mathcal{L}} \big(P_{ij,t} + P_{ij,t}^{\text{LS}}\big)+P_{i,t}^{S}\nonumber
&
\\
& & & & \ForAll{\forall\, i\in\mathcal{N},\, t\in\mathcal{T}}\quad
\IEEEyesnumber\label{eq:5} \\[-2pt]
&
Q_{i,t}^{\text{L}}
&=&
Q_{i,t}^{\text{IN}} + \sum_{ki\in\mathcal{L}} Q_{ki,t}\nonumber
&
\\
& & &
- \sum_{ij\in\mathcal{L}} \big(Q_{ij,t} + Q_{ij,t}^{\text{LS}}\big)
+ Q_{i,t}^{S}
&
\IEEEyesnumber\label{eq:6}
\\
& & & & \ForAll{\forall\, i\in\mathcal{N},\, t\in\mathcal{T}}\nonumber \\[-2pt]
&
P_{i,t}^{S}
&=&
Q_{i,t}^{S} = 0
&
\IEEEyesnumber \label{eq:7} \\[-2pt]
& & & & \ForAll{\forall\, i\neq 0,\, t\in\mathcal{T}}\nonumber \\[-2pt]
&
P_{ij,t}^{\text{LS}}
&=&
R_{ij}\,\frac{P_{ij,t}^2 + Q_{ij,t}^2}{V_{i,t}^2}
&
\IEEEyesnumber \label{eq:8}
\\
& & & & \ForAll{\forall\, (ij)\in\mathcal{L},\, t\in\mathcal{T}}\nonumber \\[-2pt]
&
Q_{ij,t}^{\text{LS}}
&=&
X_{ij}\,\frac{P_{ij,t}^2 + Q_{ij,t}^2}{V_{i,t}^2}
&
\IEEEyesnumber \label{eq:9}
\\
& & & & \ForAll{\forall\, (ij)\in\mathcal{L},\, t\in\mathcal{T}}\nonumber \\[-2pt]
&
V_{j,t}^2
&=&
V_{i,t}^2
&
\IEEEyesnumber \label{eq:10}
\\
&&&
- 2\big(R_{ij}P_{ij,t} + X_{ij}Q_{ij,t}\big)\nonumber
&
\\
&&&
+ \frac{(R_{ij}^2 + X_{ij}^2)\big(P_{ij,t}^2 + Q_{ij,t}^2\big)}{V_{i,t}^2}\nonumber
&
\\
& & & & \ForAll{\forall\, (ij)\in\mathcal{L},\, t\in\mathcal{T}}\nonumber \\[-2pt]
&
\underline{V}
&\leq&
V_{i,t}
\leq \overline{V}
&
\IEEEyesnumber \label{eq:11} \\[-2pt]
& & & & \ForAll{\forall\, i\in\mathcal{N},\, t\in\mathcal{T}}\nonumber \\[-2pt]
&
P_{ij,t}^2 &+& Q_{ij,t}^2
\leq
\bar{I}_{ij}^2\, V_{i,t}^2
&
\IEEEyesnumber \label{eq:12} \\[-2pt]
& & & & \ForAll{\forall\, (ij)\in\mathcal{L},\, t\in\mathcal{T}}\nonumber
\end{IEEEeqnarray}
\endgroup

The objective function \eqref{eq:1} contains two components: (i) the curtailment cost, computed by applying a constant coefficient $c^{\text{CU}}$ to the curtailed connection capacity, defined as the difference between the connected and allocated capacities, and (ii) the dynamic grid usage penalty cost $C^{\text{G}}(t)$, which penalizes capacity allocations that lead to operating network elements closer to their technical limits, and consequently may lead to overloading network elements in unforeseen conditions. 

Constraint~\eqref{eq:2} ensures that the total allocated capacity to all DERs does not exceed the available grid capacity at each time step. The LIFO rule is enforced in~\eqref{eq:3}. DERs are indexed according to priority, with lower indices indicating higher priority. The set $\mathcal{H}(i)$ contains all DERs with higher priority than DER $i$. Each DER $i$ is only allowed to receive a proportional allocation that does not exceed the minimum allocation ratio of any higher-priority DER. 

The dynamic grid usage penalty cost, modeled in~\eqref{eq:4}, represents the economic impact of the defined grid-usage penalty signal within the optimization objective. The formulation comprises a constant base charge, $c_0$, and a penalty component that varies with the line loading ratio $L_{ij,t}$. The penalty function, formulated in~\eqref{eq:13}, is zero when $L_{ij,t}$ remains within the safe limit $L_{\text{S}}$, but increases quadratically once this threshold is exceeded, thereby discouraging operation of network elements close to their technical limits. 
\begin{equation}
\begin{aligned}
    f(L_{ij,t}) &=
    \begin{cases}
        0, & L_{ij,t} \leq L_{\text{S}} \\
        \gamma (L_{ij,t} - L_{\text{S}})^2, & L_{\text{S}} < L_{ij,t} \leq 1
    \end{cases}
    \label{eq:13} \\
    &\hspace{12em} \forall\, ij \in \mathcal{L},\; t \in \mathcal{T}
\end{aligned}
\end{equation}

\hspace{1.5em}\text{where, } $L_{ij,t} = \dfrac{I_{ij,t}}{\bar{I}_{ij}}$.

Finally, the OPF constraints in \eqref{eq:5}--\eqref{eq:12}~\cite{ref23} guarantee that the injected power by the DERs comply with all physical and operational constraints and limitations of the distribution network. Expressions~\eqref{eq:5}–\eqref{eq:6} enforce active and reactive power balance at each node, including line losses, DER power injections (generation), loads and line flows. Since the proposed allocation model focuses on generation limits, the active and reactive load demands are treated as fixed day-ahead parameters. Slack bus injections are defined in Eq.~\eqref{eq:7}, while line losses are modeled in Eqs.~\eqref{eq:8}–\eqref{eq:9}. Voltage magnitude drops due to power flow and line impedance are captured by Eq.~\eqref{eq:10}, where \(R_{ij}\) and \(X_{ij}\) represent the resistance and reactance of line \((i,j)\), respectively. Finally, nodal voltage limits are enforced by Constraint~\eqref{eq:11}, and Eq.~\eqref{eq:12} defines the relationship used to compute line current magnitudes.

\subsubsection{Lower-Level Problem (DER perspective)}
The lower-level problem captures the decision-making process of each DER owner. The objective function in~\eqref{eq:LL_obj_clean} maximizes the total profit of individual DERs by accounting for both revenues and operational costs. In this formulation, the profit is computed as the revenue from actual electricity sales, valued at the electricity price $\lambda_t$, minus the operating cost represented 
by the fuel-equivalent marginal cost $\phi_{i,t}$, which is zero for renewable-based DERs. Constraint~\eqref{eq:LL_constraints} ensures that each DER’s actual power injection 
remains within its technical bounds and the allocated power limit during the scheduling horizon.
\begin{IEEEeqnarray}{l}
\hspace{-1.2em}\max_{\{ P_{i,t}^{\text{IN}} \}_{i\in\mathcal{N},\,t\in\mathcal{T}}}
\;\sum_{t \in \mathcal{T}}
\Big[ (\lambda_t - \phi_{i,t})\, P_{i,t}^{\text{IN}} \Big]\Delta t
\IEEEyesnumber\label{eq:LL_obj_clean}
\end{IEEEeqnarray}
\begin{IEEEeqnarray}{r rCl l}
\IEEEeqnarraymulticol{5}{l}{\text{s.t.}}\nonumber\\
&
\underline{P}_{i}^{\text{IN}}
&\leq&
P_{i,t}^{\text{IN}}
\;\le\;
\min\!\left\{ \overline{P}_{i}^{\text{IN}},\; P_{i,t}^{\text{A}} \right\}
& \ForAll{\forall\, i \in \mathcal{N},\, t \in \mathcal{T}}
\IEEEyesnumber\label{eq:LL_constraints}
\end{IEEEeqnarray}
\subsection{Solution Approach}
\label{subsec:acopf-linearization}
The solution approach consists of two main steps. Since the original formulation involves a nonlinear AC optimal power flow (OPF) problem, directly solving it would be computationally challenging. Therefore, in the first step, the AC OPF formulation is linearized using the LinDistFlow approximation~\cite{ref24} to improve computational tractability. In the second step, the bilevel problem is reformulated as a single-level optimization problem by replacing the lower-level problem with its Karush–Kuhn–Tucker (KKT) conditions~\cite{ref27}. The details of these steps are provided below.

\subsubsection{AC OPF Linearization Using LinDistFlow Approximation}
The AC OPF formulation in Eqs. \eqref{eq:5}--\eqref{eq:12} includes nonlinear constraints arising from the modeling of active and reactive power losses, voltage drops across lines, and line current magnitudes. 
To improve tractability, the LinDistFlow approximation introduced in~\cite{ref24} is adopted to replace these nonlinear constraints with linear expressions. This approximation is commonly used in radial distribution networks because it provides good accuracy while substantially reducing computational complexity~\cite{ref25}. LinDistFlow relies on the following assumptions: (i) small voltage angle differences, (ii) limited voltage deviations from nominal, and (iii) relatively small line losses. 

Applying these assumptions, the OPF constraints are linearized by omitting loss terms in the power balance equations, removing quadratic components in the voltage expressions, defining $U_{i,t}=V_{i,t}^2$, and approximating thermal limits through a conservative $\ell_1$-norm representation. Absolute values in the current constraints are handled via auxiliary variables, and using $\underline{V}$ ensures conservativeness across the feasible voltage range~\cite{ref26}. The resulting linear formulation is given in Eqs.  \eqref{eq:16}--\eqref{eq:22}, which replace Eqs. \eqref{eq:5}--\eqref{eq:12} in the upper level problem:
\begin{align}
P_{i,t}^{\text{L}} = P_{i,t}^{\text{IN}} + \sum_{ki \in \mathcal{L}} P_{ki,t} - \sum_{ij \in \mathcal{L}} P_{ij,t} + P_{i,t}^{S},  \ForAll{\forall\, i \in \mathcal{N},\, t \in \mathcal{T}}
\label{eq:16}\\
Q_{i,t}^{\text{L}} = Q_{i,t}^{\text{IN}}+\sum_{ki \in \mathcal{L}} Q_{ki,t} - \sum_{ij \in \mathcal{L}} Q_{ij,t} + Q_{i,t}^{S}, \ForAll{\forall\, i \in \mathcal{N},\, t \in \mathcal{T}}
\label{eq:17}\\
P_{i,t}^{S} = Q_{i,t}^{S} = 0, 
\quad \quad\quad\quad \quad\quad \quad\quad \quad \ForAll{\forall\, i\neq0,\, t \in \mathcal{T}}
\label{eq:18}\\
P_{ij,t}^{\text{LS}} = Q_{ij,t}^{\text{LS}} = 0,
\quad \quad \quad \quad\quad\quad\quad\quad \ForAll{\forall\, ij \in \mathcal{L},\, t \in \mathcal{T}}
\label{eq:19}\\
U_{j,t} = U_{i,t} - 2\big(R_{ij} P_{ij,t} + X_{ij} Q_{ij,t}\big), 
 \ForAll{\forall\, ij \in \mathcal{L},\, t \in \mathcal{T}}
\label{eq:20}\\
\underline{V}^{2} \leq U_{i,t} \leq \overline{V}^{2}, 
\quad\quad\quad\quad\quad \quad \quad \quad \quad \ForAll{\forall\, i \in \mathcal{N},\, t \in \mathcal{T}}
\label{eq:21}\\
|P_{ij,t}| + |Q_{ij,t}| \leq \bar{I}_{ij}\,\underline{V}, 
\quad\quad\quad\quad \quad \quad \ForAll{\forall\, ij \in \mathcal{L},\, t \in \mathcal{T}}
\label{eq:22}
\end{align}

\subsubsection{KKT Conditions to Reformulate the Bilevel Problem into a Single-Level Problem}
Following standard academic practice, the lower level problem is replaced with its KKT conditions~\cite{ref27} to reformulate the problem as a single-level model. Since the lower-level problem is convex (linear objective and constraints), the KKT conditions are both necessary and sufficient for optimality. The complementary slackness constraints are embedded directly into the model as relaxed bilinear constraints using a small tolerance parameter $\epsilon$.
\subsection{KPIs Definition}
\label{sec:KPIs Definition}
A set of KPIs is introduced to capture the key aspects of the capacity allocation problem from both the DSO and DER operators’ perspectives. These include grid utilization, DER curtailment, allocation variability, and economic viability. The definition and formulation of each KPI are detailed below.

\subsubsection{Grid Utilization}
This KPI quantifies how efficiently the available network capacity is utilized. It is assessed using line utilization levels, which reflect how effectively the proposed strategy accommodates DER integration without violating thermal limits. The normalized line loading is defined as:
   \begin{equation}
    L_{ij,t} = \frac{I_{ij,t}}{\bar{I}_{ij}},
    \quad \forall\, (ij)\in\mathcal{L},\ \forall\, t\in\mathcal{T}
    \label{eq:33}
    \end{equation}

 \subsubsection{DER Curtailment} It represents the deviation between the allocated and connected capacity profiles over time. It quantifies the portion of the connected capacity that cannot be delivered due to network constraints and the allocation rules considered. 
The following metrics are used to evaluate the level of DER curtailment:
\setlength{\itemsep}{0pt}
\setlength{\parsep}{0pt}

        \begin{itemize}
        \item[] \textbf{Curtailment Severity}, the average size of curtailment events for each DER:
        \begin{equation}
        \bar{P}^{\text{curt}}_i
        = \frac{1}{|\mathcal{T}|}
        \sum_{t\in\mathcal{T}}
        \left( P^{\mathrm{C}}_{i,t} - P^{\mathrm{A}}_{i,t} \right),
        \quad \forall\, i\in\mathcal{N}
        \label{eq:29}
        \end{equation}
        \item[] \textbf{Total Curtailed Capacity}, defined as the cumulative reduction in allocated connection capacity across all DERs over the scheduling horizon:
         \begin{equation}
        \Gamma^{\text{curt}}_{\text{total}}
        = \sum_{i\in\mathcal{N}} \sum_{t\in\mathcal{T}}
        \left( P^{\mathrm{C}}_{i,t} - P^{\mathrm{A}}_{i,t} \right)\ 
        \label{eq:30_total}
        \end{equation}
    
    \item[] \textbf{Number of Curtailment Intervals}, indicating the number of time intervals during which DER $i$ is curtailed over the scheduling horizon:
    \begin{equation}
    N^{\text{curt}}_i
    =
    \sum_{t \in \mathcal{T}}
    \mathbb{1}_{\left\{ P^{\mathrm{A}}_{i,t} < P^{\mathrm{C}}_{i,t} \right\}},
    \quad \forall\, i \in \mathcal{N}
    \label{eq:32}
    \end{equation}
    where, $\mathbb{1}_{\{\cdot\}}$ denotes the indicator function, which equals 1 if the condition is satisfied and 0 otherwise.
    \end{itemize}
\subsubsection{Allocation Variability}
\label{subsec:allocation_consistency}

This is a key consideration for DER owners, as high variability in allocated connection capacity can be difficult to forecast or manage, which in turn directly affects their ability to plan generation and participation in electricity markets. This can lead to operational inefficiencies and reduced economic benefits. Therefore, allocation variability is assessed in terms of the temporal stability and consistency of the allocated capacity over the scheduling horizon. To capture these aspects, two complementary metrics are defined.
\begin{itemize}
\item[] \textbf{Temporal Stability} This metric evaluates short-term fluctuations in allocated capacity between successive time steps. It is quantified through temporal variability, defined as the average normalized change in allocated power limits between consecutive time intervals. Lower values indicate higher temporal stability.
\begin{equation}
\overline{\Delta P}^{\mathrm{A}}_{i}
=
\frac{1}{|\mathcal{T}|-1}
\sum_{t=2}^{|\mathcal{T}|}
\frac{
\left| \Delta P^{\mathrm{A}}_{i,t} \right|
}{
\max\left(P^{\mathrm{C}}_{i,t-1}, P^{\mathrm{C}}_{i,t}\right)
},
\quad \forall\, i\in\mathcal{N}
\end{equation}
with 
\begin{equation}
\Delta P^{\mathrm{A}}_{i,t}
=
P^{\mathrm{A}}_{i,t} - P^{\mathrm{A}}_{i,t-1},
\quad \forall\, i \in \mathcal{N},\ \forall\, t \geq 2
\end{equation}
\item[] \textbf{Allocation Consistency} This metric captures the variability of normalized allocations over time. It is quantified as the standard deviation of the normalized allocated power limits across the time horizon. Lower values indicate more consistent allocation profiles:
\begin{equation}
r_{i,t}
=
\frac{P^{\mathrm{A}}_{i,t}}{P^{\mathrm{C}}_{i,t}},
\quad \forall\, i \in \mathcal{N},\ \forall\, t \in \mathcal{T}
\end{equation}
\begin{equation}
\sigma_i
=
\sqrt{
\frac{1}{|\mathcal{T}|}
\sum_{t \in \mathcal{T}}
\left(
r_{i,t} - \bar{r}_i
\right)^2
},
\quad \forall\, i \in \mathcal{N}
\end{equation}
\end{itemize}
\subsubsection{Economic Viability}
Evaluating economic considerations is essential for both the DSO and DER owners, as financial incentives strongly influence participation in new contractual schemes. Moreover, a DER owner may choose not to inject its available production if operational costs outweigh the expected revenue. To capture these aspects, two metrics are defined:
\begin{itemize}
    \item[] \textbf{Curtailment Cost}, represented by the second term in the objective function of the upper-level problem in Eq.~\eqref{eq:1}, is given by \(c^{\text{CU}} (P_{i,t}^{\text{C}} - P_{i,t}^{\text{A}})\). This term represents a system-level penalty associated with unutilized DER generation and is introduced to discourage excessive curtailment from the DSO perspective.
    \item[] \textbf{DER Revenue}, defined by the lower-level objective function in Eq.~\eqref{eq:LL_obj_clean}, is given by \((\lambda_t - \phi_{i,t})\, P_{i,t}^{\text{IN}}\). It represents the net income earned by DERs from power injections, accounting for electricity market prices and DER-specific operational costs. It captures the economic incentive for DERs to maximize power injection under the allocated network capacity. 
\end{itemize}
\FloatBarrier
\subsection{Benchmark Models}
\label{app:BENCHMARK}
Two benchmark models are considered to evaluate the proposed bilevel allocation model, namely the LIFO-only and OPF-only allocation models. Both models employ the curtailment cost minimization objective of the upper-level problem, while differing in the set of enforced constraints.
\subsubsection{LIFO-only benchmark model} \label{subsec:LIFO_benchmark}
This benchmark minimizes total DER curtailment using the first term of the upper-level objective reformulated in~\eqref{eq:31}. The model enforces the total grid capacity limit given by~\eqref{eq:2} and the LIFO constraint formulated in~\eqref{eq:3}, while excluding the OPF constraints. Therefore, this benchmark represents a capacity allocation model that is not aware of network operating constraints and does not explicitly ensure grid feasibility.
\begin{equation} \min_{\{ P_{i,t}^{\mathrm{A}} \}} c^{\text{CU}}\sum_{t \in \mathcal{T}} \sum_{i \in \mathcal{N}} \left( P_{i,t}^{\mathrm{C}} - P_{i,t}^{\mathrm{A}} \right). \label{eq:31} \end{equation} 
\indent Since an allocation strategy that violates network operating constraints cannot be considered practically valid, an ex post solution feasibility check is performed by running a power flow (PF) analysis using the obtained allocation limits. If infeasibilities are detected, the maximum grid capacity available to DERs, together with the connected load demand capacities, is iteratively reduced until a grid-feasible solution is achieved while preserving the LIFO-based allocation structure. 
\subsubsection{OPF-only benchmark model} \label{subsec:OPF_benchmark} This benchmark adopts the same curtailment cost minimization objective as~\eqref{eq:31}. The model enforces the total grid capacity limit in~\eqref{eq:2} together with the full set of OPF constraints~\eqref{eq:5}--\eqref{eq:12}, while excluding the LIFO constraint~\eqref{eq:3}. Consequently, the obtained allocations are driven solely by network feasibility requirements, representing a DSO-centric allocation model that does not explicitly account for DER-level decision-making or economic objectives.
\section{Case Study}
\begin{figure}[t]
\centering
\includegraphics[width=0.9\linewidth]{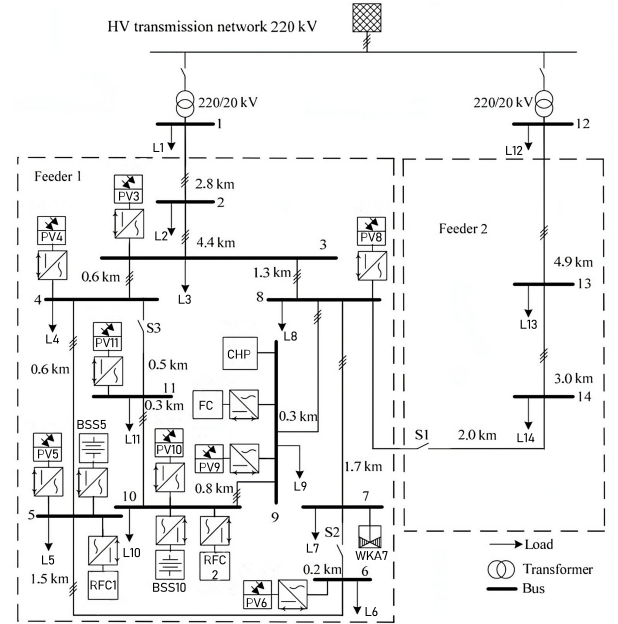}
\caption{CIGRE MV distribution network with connected loads and DERs.}
\label{fig:casestudy}
\end{figure}

The proposed model is evaluated on the CIGRE MV distribution network with various connected DER units, including eight photovoltaic (PV) systems, one wind turbine (WKA), one combined heat and power (CHP) unit, one fuel cell (FC) and two residential fuel cell (RFC) units \cite{ref28} (Fig.~\ref{fig:casestudy}). In accordance with the adopted LIFO-based prioritization rule and to align the study with sustainability objectives, higher allocation priority is assigned to RES-based DER units, including PV and wind generation. To induce congested operating conditions, the connected capacities of loads and DERs are proportionally increased, with the resulting capacities summarized in Table~\ref{tab:load_der_capacities}. Loads are considered fixed parameters and the proposed capacity allocation model is applied exclusively to DER generation units. The time-series data for load demand and DER generation are generated by multiplying the connected capacities of loads and DERs in Table~\ref{tab:load_der_capacities} with normalized 24-hour profiles obtained from \cite{ref29}, \cite{ref30}, and \cite{ref31}, corresponding to load demand, RES, and CHP generation. The resulting profiles correspond to a representative day reflecting typical average operating conditions to evaluate the proposed model under congested network conditions. The bilevel capacity allocation model is implemented in Python using the Pyomo environment and solved with the Gurobi solver. 
The effectiveness of the proposed bilevel capacity allocation model is evaluated by assessing the defined KPIs and, where applicable, comparing them with those obtained from the two benchmark models.
\begin{table}[t]
\centering
\small
\caption{Load and DER connected capacities}
\label{tab:load_der_capacities}
\renewcommand{\arraystretch}{0.85}
\setlength{\tabcolsep}{3.5pt}

\begin{tabular}{c c r @{\hspace{10pt}} c l r}
\toprule
\multicolumn{3}{c}{\textbf{Load}} &
\multicolumn{3}{c}{\textbf{DER}} \\
\cmidrule(lr){1-3} \cmidrule(lr){4-6}
\textbf{Bus} & \textbf{Name} & $\mathbf{S^{\mathrm{L}}}$(MVA) &
\textbf{Bus} & \textbf{Name} & $\mathbf{S^{\mathrm{C}}}$(MVA) \\
\midrule
1  & $L_1$    & 40.80 & 3  & PV3  & 4.00 \\
3  & $L_3$    & 1.10  & 4  & PV4  & 4.00 \\
4  & $L_4$    & 0.89  & 5  & PV5  & 6.00 \\
5  & $L_5$    & 1.50  & 5  & RFC1 & 10.00 \\
6  & $L_6$    & 1.13  & 6  & PV6  & 6.00 \\
7  & $L_7$    & 0.18  & 7  & WKA7 & 7.50 \\
8  & $L_8$    & 1.21  & 8  & PV8  & 6.00 \\
9  & $L_9$    & 1.35  & 9  & PV9  & 6.00 \\
10 & $L_{10}$ & 1.14  & 9  & CHP  & 10.00 \\
11 & $L_{11}$ & 0.68  & 9  & FC   & 10.00 \\
12 & $L_{12}$ & 41.16 & 10 & PV10 & 8.00 \\
13 & $L_{13}$ & 0.08  & 10 & RFC2 & 10.00 \\
14 & $L_{14}$ & 1.21  & 11 & PV11 & 2.00 \\
\bottomrule
\end{tabular}
\end{table}
\subsection{Grid Utilization}
\begin{figure}[t]
\centering
\includegraphics[width=0.9\linewidth]{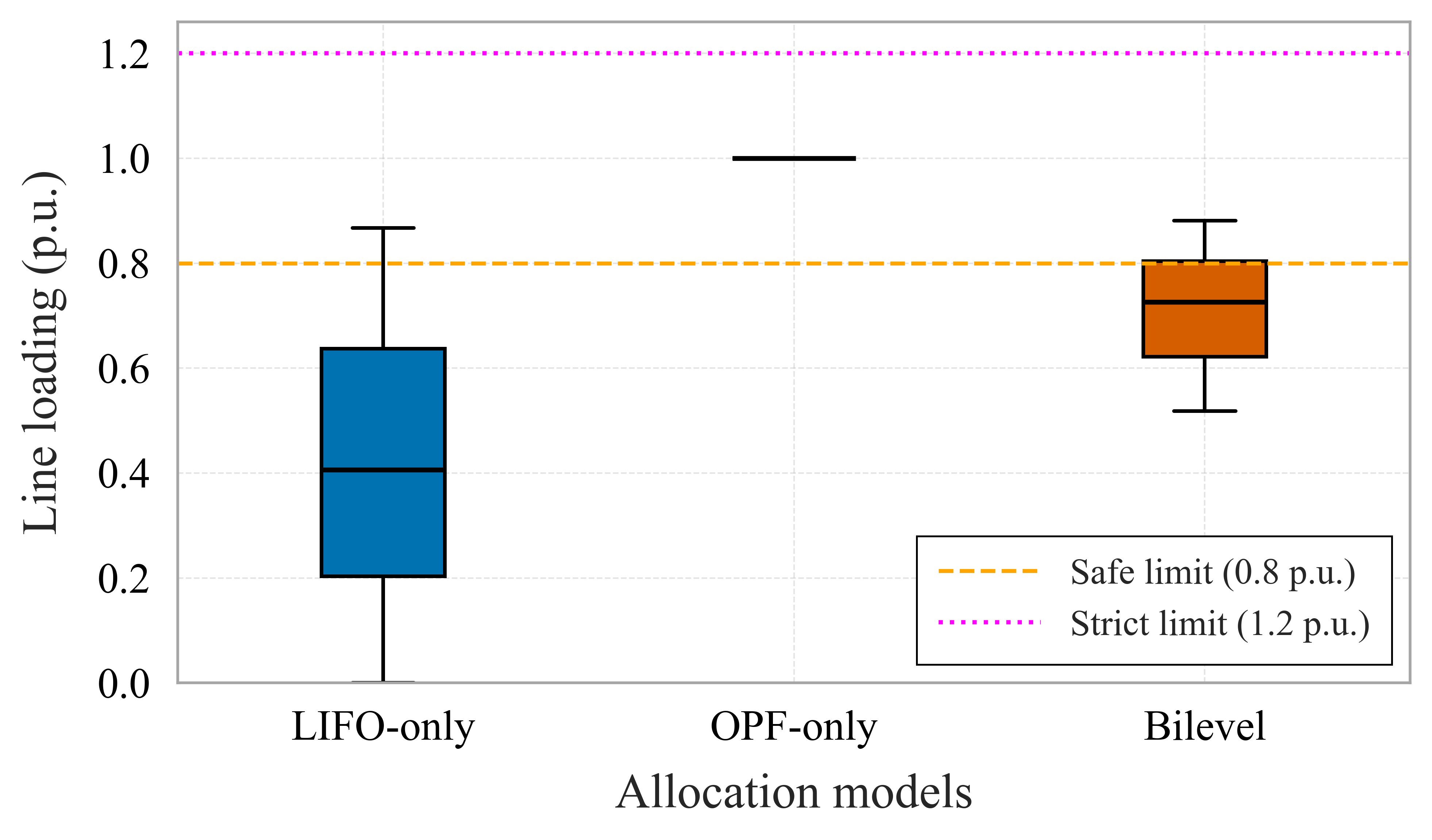}
\caption{Line loading distributions across allocation models, with min–max whiskers and operational loading limits.}
\label{fig:line_loading_profile}
\end{figure}

Figure~\ref{fig:line_loading_profile} compares the distributions of normalized line loading across all network lines for the LIFO-only, OPF-only, and bilevel allocation models, together with the safe (0.8 p.u.) and strict (1.2 p.u.) loading limits. The boxplots show the median, interquartile range, and minimum/maximum loading levels across the network lines.

The LIFO-only model exhibits a wide spread of loading levels, with most lines operating below the safe loading limit. This results from the ex post solution checker, which iteratively reduces the allocable DER capacity to restore grid feasibility while preserving the LIFO-based allocation structure, leading to conservative and uneven network utilization. In contrast, the OPF-only model drives all lines close to 1.0 p.u. loading with almost no variation, as the curtailment minimization objective leads the optimization to utilize the available network capacity up to the operational loading limit enforced by the OPF constraints. The bilevel model balances these two extremes by increasing network utilization relative to the LIFO-only model while avoiding uniformly high loading levels, resulting in a more efficient and operationally robust utilization of network capacity.
\begin{figure}[t]
\centering
\includegraphics[width=0.9\linewidth]{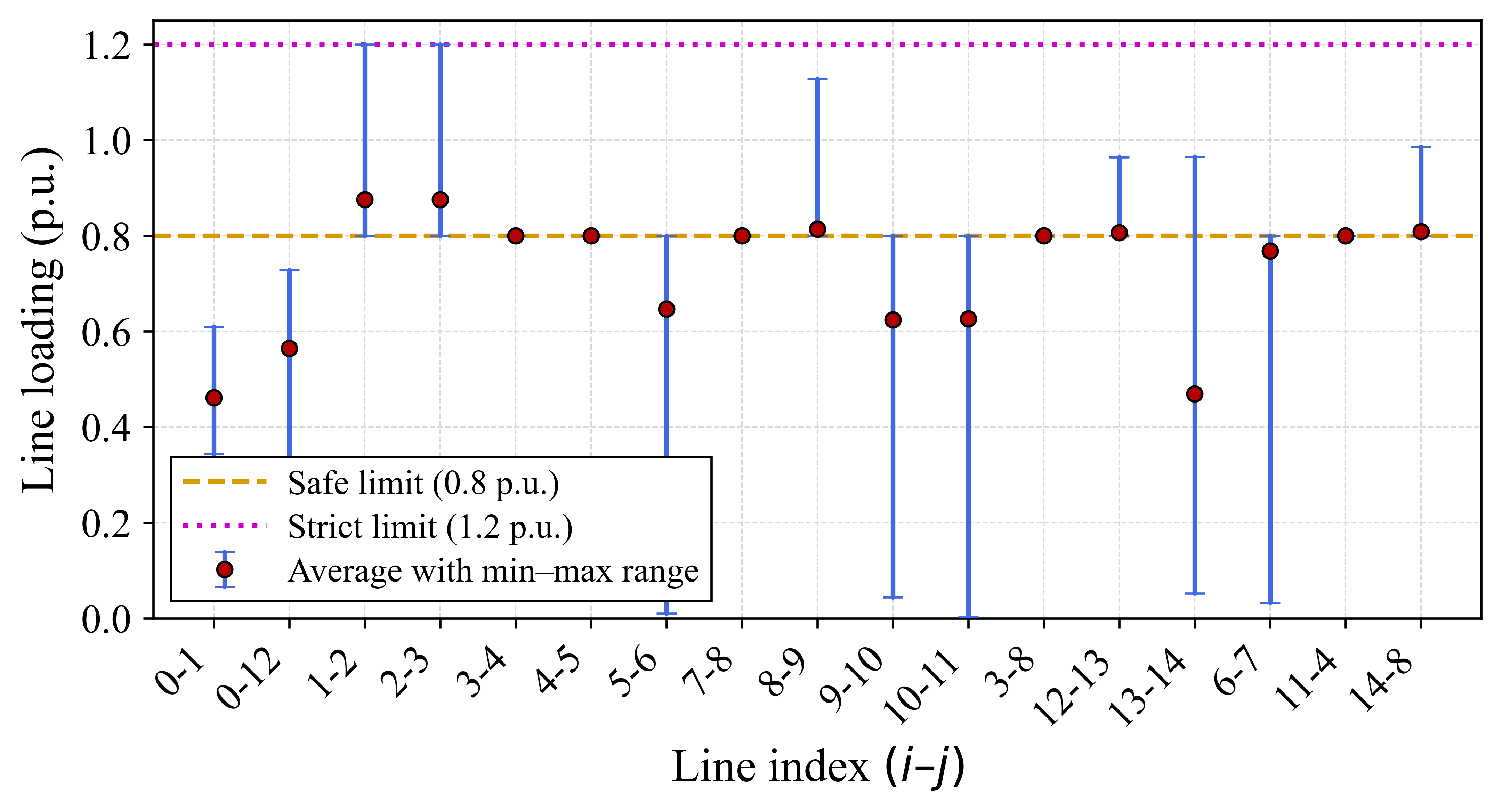}
\caption{Line loading across network lines in the bilevel allocation model.}
\label{fig:line_loading_profile_bilevel}
\end{figure}
Figure~\ref{fig:line_loading_profile_bilevel} presents a line-level assessment of grid utilization under the proposed bilevel allocation model, showing the range of line loading across all lines over the analyzed time horizon. The results indicate that most lines operate below the safe loading threshold of 0.8~p.u., while all remain strictly within the thermal limit of 1.2~p.u., ensuring secure and feasible network operation. A subset of lines, namely (1–2), (2–3), (8–9), (12–13), (13–14), and (14–8), exhibits higher average loading together with elevated maximum loading levels, indicating periods of increased stress. These lines are either connected to the main substation or located near regions with high DER penetration, where aggregated generation leads to inherently higher utilization due to the network topology (Fig.~\ref{fig:casestudy}).
\subsection{DER Curtailment}
\begin{figure}[t]
\centering
\includegraphics[width=0.9\linewidth]{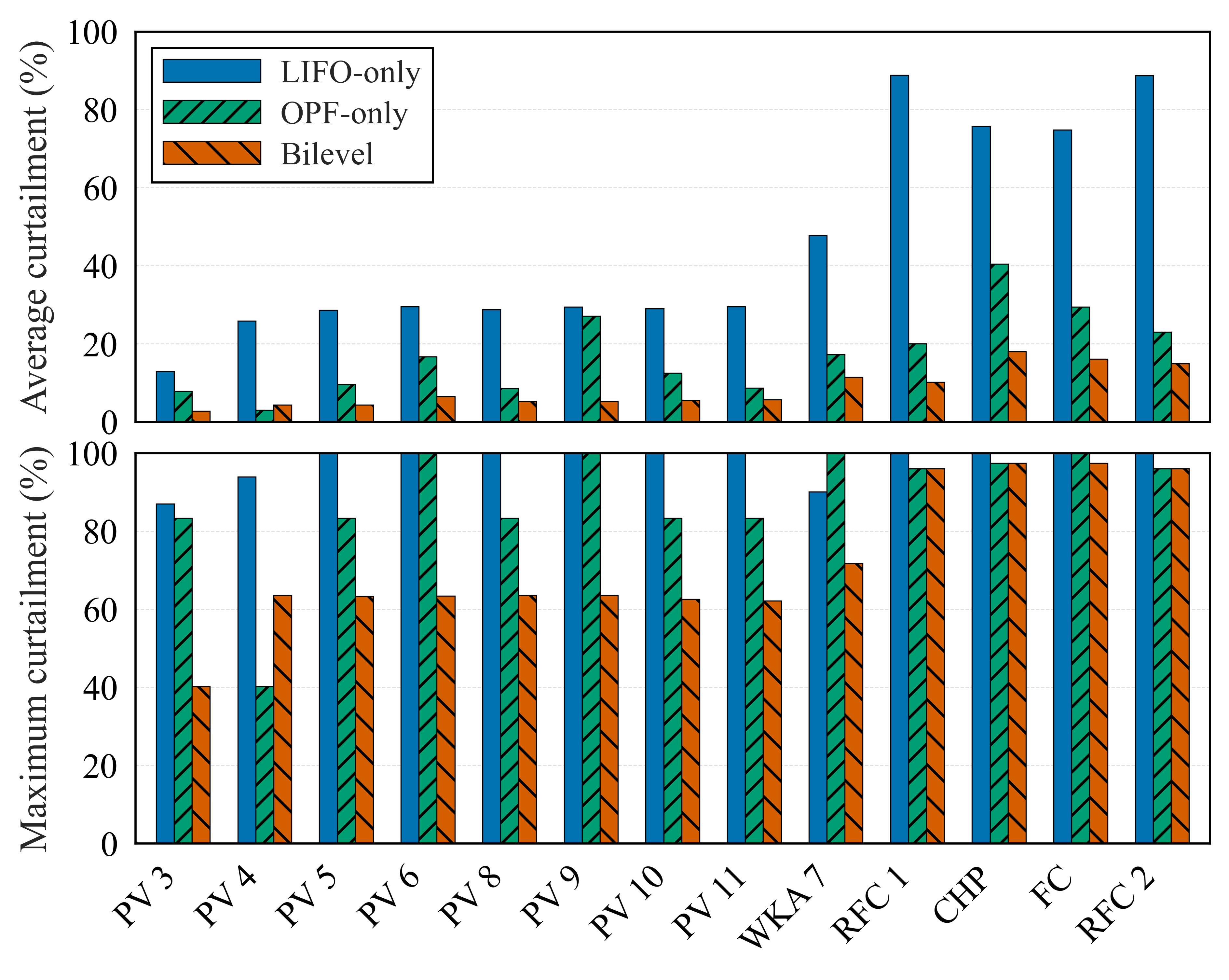}
\caption{Average and maximum curtailment severity per DER across three allocation models. Top: average curtailment as a percentage of the connected capacity, bottom: the worst-case (maximum) curtailment experienced by each DER over the 24-h horizon.}
\label{fig:curtailment_severity}
\end{figure}

Figure~\ref{fig:curtailment_severity} illustrates the average and maximum curtailment severity for each DER under the three allocation models, with DERs ordered from left to right according to decreasing priority. The LIFO-only allocation model consistently results in the highest curtailment levels, particularly for lower-priority DERs. While higher-priority units exhibit moderate average curtailment levels (typically below 30\%), lower-priority DERs experience significantly higher values, reaching 70--90\% in some cases, and maximum curtailment approaches 100\%. This behavior reflects the strict sequential nature of LIFO, in which higher-priority DERs are always served first and lower-priority units bear most of the curtailment required to relieve network congestion. In contrast, the OPF-only allocation model substantially reduces average curtailment across most DERs, generally keeping it below 40\%. However, the maximum curtailment experienced by individual DERs remains high during congested periods, with worst-case values exceeding 80\%. This reflects that the OPF-only model primarily minimizes total system curtailment without explicitly accounting for allocation order at the individual DER level.

The bilevel allocation model achieves the lowest average curtailment across almost all DERs while also mitigating extreme curtailment events. In particular, the average curtailment is generally around or below 20\%, representing a reduction of approximately 10--20 percentage points (p.p.) compared to the OPF-only model. Compared to the LIFO-only model, the reduction is modest for higher-priority DERs but becomes substantial for lower-priority DERs, reaching up to 50--70 p.p. in some cases. Moreover, while maximum curtailment under LIFO frequently reaches or approaches 100\%, the bilevel model results in significantly lower worst-case curtailment levels, remaining below 70\%.


\begin{figure}[t]
\centering
\includegraphics[width=0.9\linewidth]{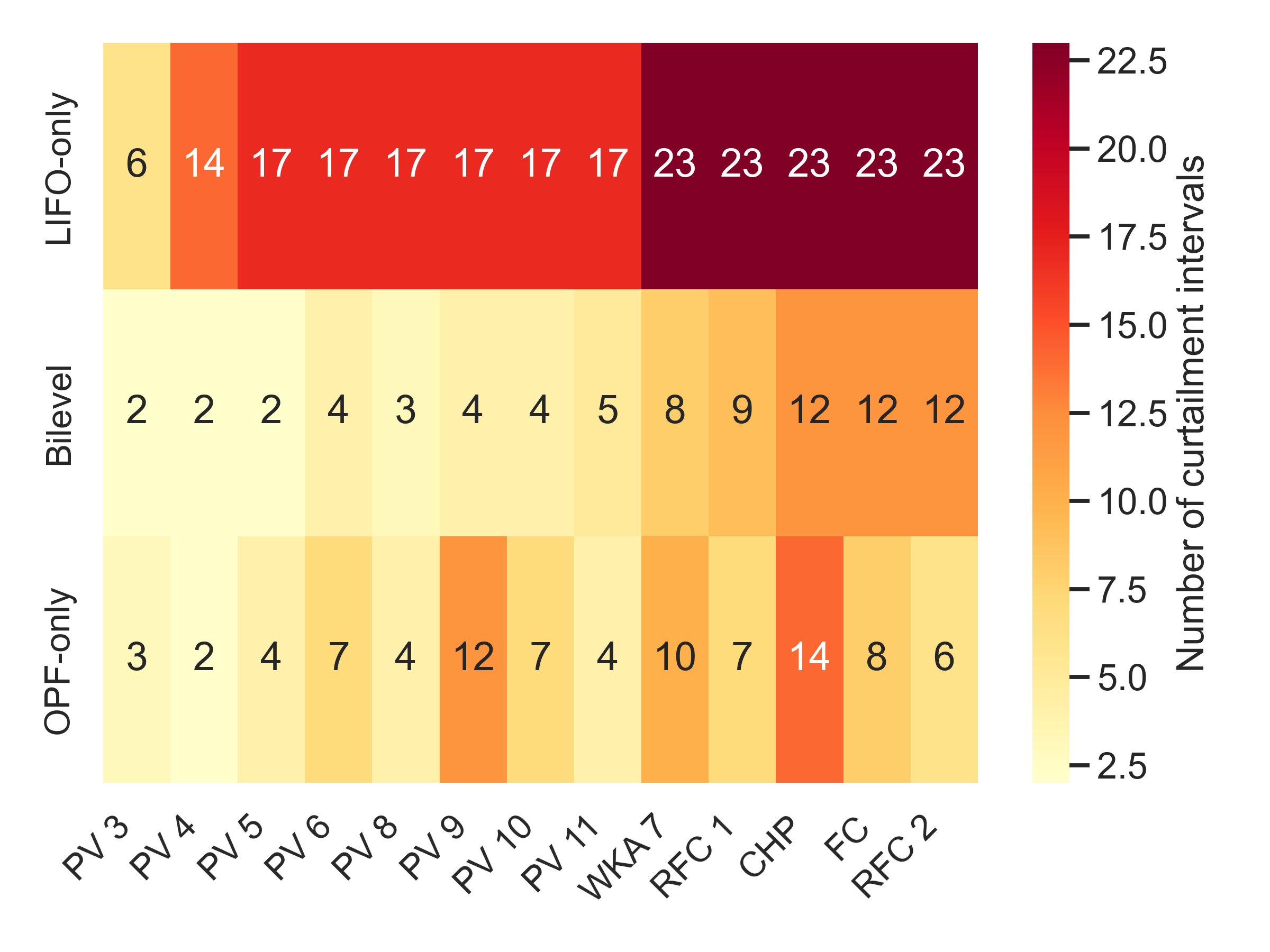}
\caption{Number of curtailment intervals per DER across allocation models.}
\label{fig:curtailment_frequency}
\end{figure}

Figure~\ref{fig:curtailment_frequency} compares the number of curtailment intervals experienced by each DER under the three allocation models over the 24-h horizon. The LIFO-only strategy results in the highest number of curtailment intervals, with curtailment strongly concentrated on lower-priority DERs. High-priority units such as PV~3 and PV~4 are curtailed relatively few times (6 and 14 intervals, respectively), whereas most lower-priority DERs are curtailed for up to 23 intervals. This behavior originates from the iterative ex-post feasibility checking process, where allocable capacity is progressively reduced until a feasible operating point is achieved, leading to conservative curtailment decisions.

The OPF-only strategy reduces the overall number of curtailment intervals, with values ranging from 2 to 14 intervals across DERs. However, the distribution of curtailment intervals across DERs is irregular, indicating that the OPF-only model dynamically reallocates curtailment based on network feasibility without preserving priority order or temporal stability at the DER level. In contrast, the bilevel strategy addresses both limitations simultaneously. First, it significantly decreases the number of curtailment intervals across all DERs, with high-priority units curtailed only 2--5 times and lower-priority DERs limited to at most 12 intervals. Second, it achieves a more balanced distribution of curtailment intervals among DERs, avoiding both the excessive concentration of curtailment on lower-priority units observed under LIFO and the irregular, non-predictable curtailment patterns characteristic of the OPF-only model.
\subsection{Allocation Variability}
\label{allocation_consistency}
\begin{figure}[t]
\centering
\includegraphics[width=\linewidth]{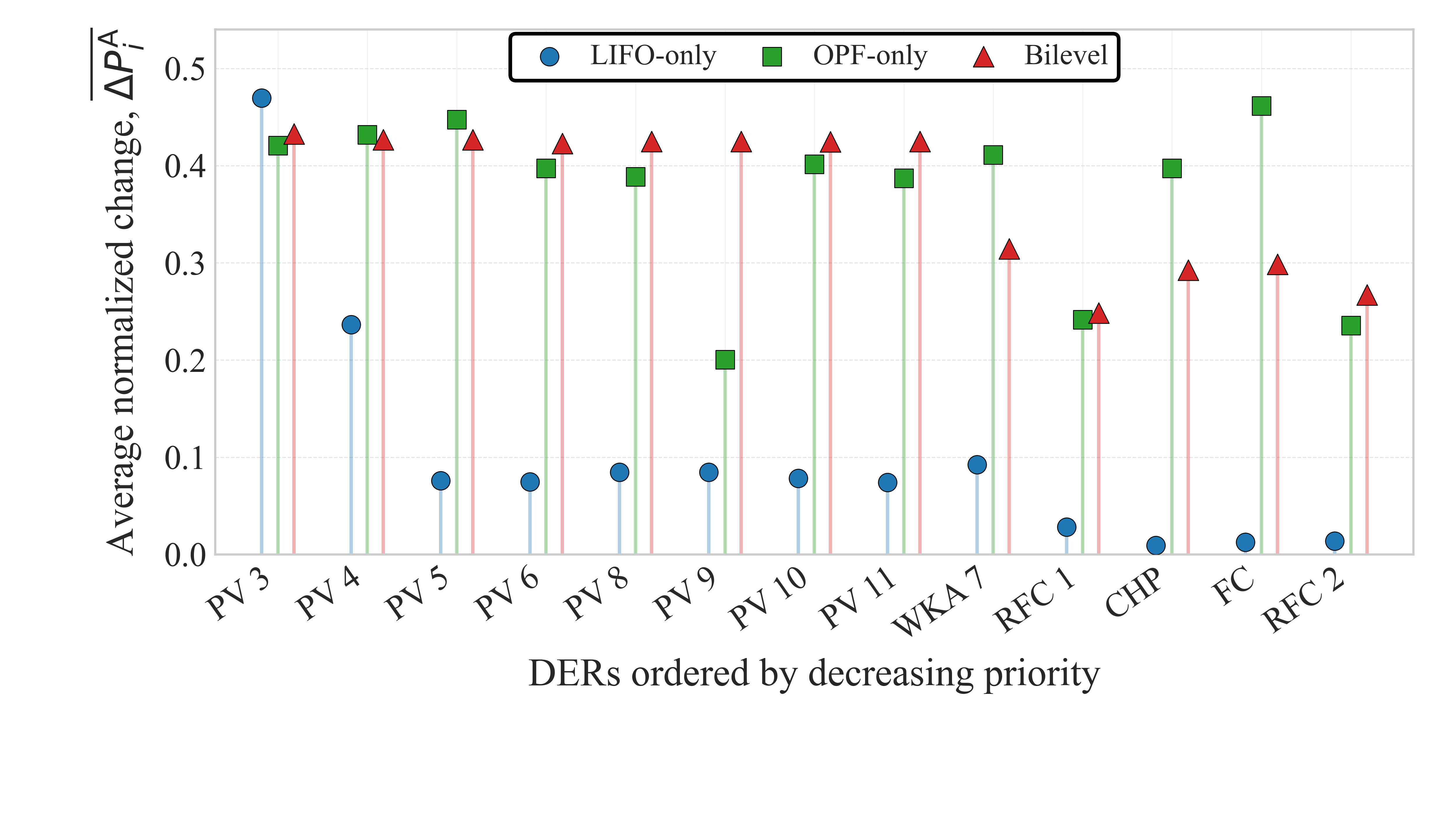}
\caption{Comparison of average normalized temporal variability in allocated capacity to DERs between consecutive time intervals across  allocation models. }
\label{fig:temporal_stability}
\end{figure}

Temporal variability of allocations is illustrated in Fig.~\ref{fig:temporal_stability}, which shows the average normalized change in allocated capacity for each DER between two successive time steps for the three allocation methods. The LIFO-only strategy exhibits the lowest temporal variability for most DERs, particularly for lower-priority units, since these DERs are frequently assigned low or zero allocations over time. In the other hand, high-priority DERs experience larger variability because the LIFO strategy, together with its ex-post feasibility checker leads to abrupt changes whenever additional capacity reductions are required to satisfy network constraints. The OPF-only strategy results in significantly higher temporal variability across almost all DERs, reflecting frequent allocation adjustments to maintain network feasibility. Moreover, the variability pattern shows no clear trend with respect to DER priority, indicating irregular and less predictable allocation dynamics. In contrast, the bilevel strategy achieves a more balanced and structured temporal behavior. Although the temporal variability remains higher than in the LIFO-only model, it follows a more consistent trend across DER priorities, with progressively lower variability for lower-priority DERs. Compared to the OPF-only strategy, the bilevel approach avoids abrupt and irregular fluctuations while providing a smoother and more uniform distribution of temporal variability among DERs, indicating improved stability in the temporal evolution of allocations across the network.

\begin{figure}[t]
\centering
\includegraphics[width=0.9\linewidth]{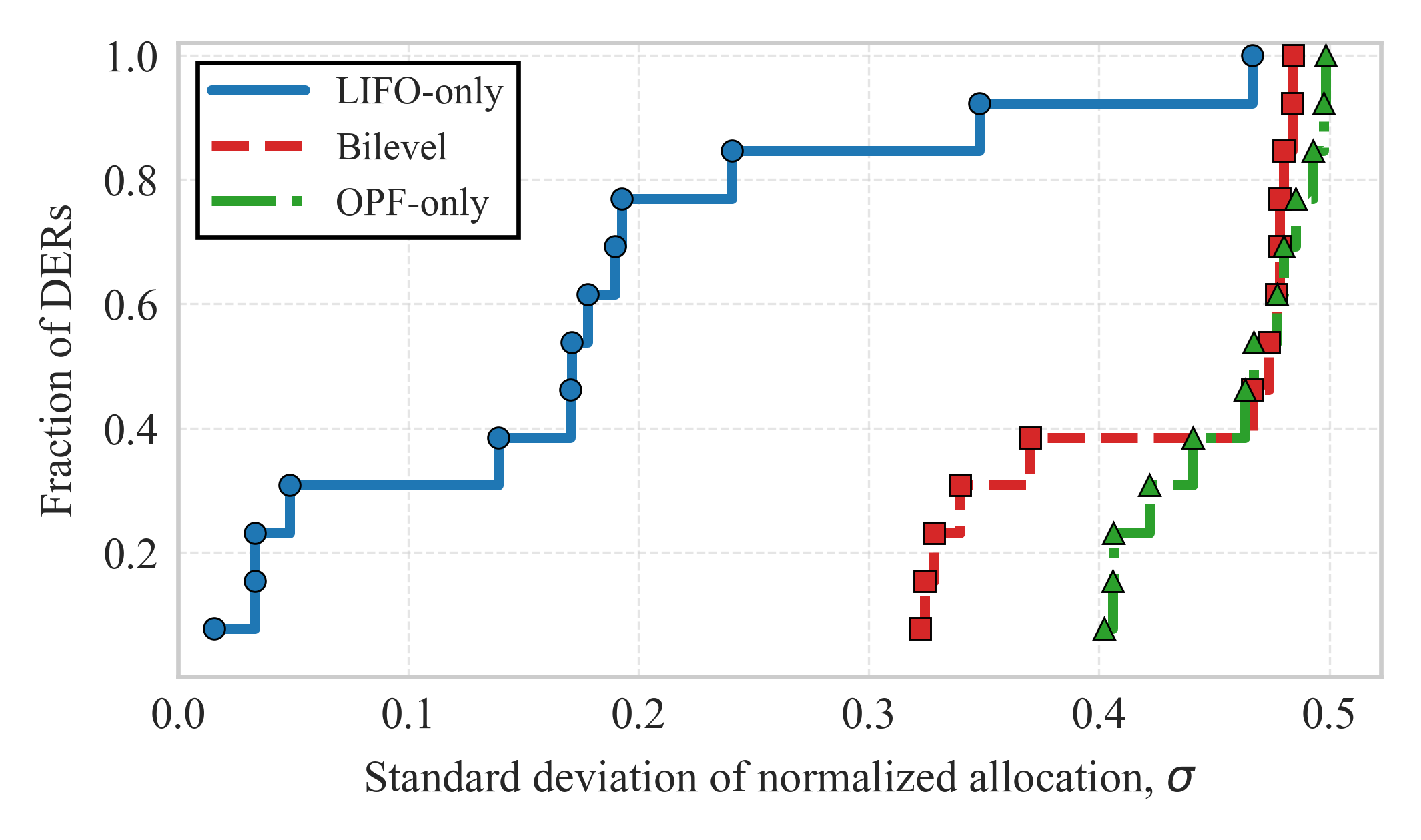}
\caption{Comparison of allocation consistency across DERs under different allocation models using ECDFs.}
\label{fig:allocation_consistency}
\end{figure}

Figure~\ref{fig:allocation_consistency} illustrates the empirical cumulative distribution functions (ECDFs) of the standard deviation of normalized allocations. The horizontal axis represents the standard deviation of allocations, while the vertical axis indicates the fraction of DERs whose standard deviation does not exceed a given value. Curves further to the left correspond to more consistent allocations, as they indicate smaller deviations from the mean allocation.

The LIFO-only strategy exhibits the lowest allocation variability for a large fraction of DERs, with approximately 70\% of DERs experiencing $\sigma \lesssim 0.20$. However, the gradual increase of the ECDF indicates heterogeneity across DERs, with higher-priority units experiencing higher variability. In contrast, the OPF-only strategy results in consistently higher variability, with most DERs exhibiting $\sigma \gtrsim 0.40$. The bilevel model achieves a more balanced performance, reducing variability compared to OPF-only, with most DERs exhibiting standard deviation values in the range of approximately $\sigma \approx 0.30\text{--}0.45$. The bilevel model achieves a balanced performance, reducing variability compared to OPF-only, with most DERs exhibiting $\sigma \approx 0.30\text{--}0.45$, while maintaining a more uniform distribution of consistency across DERs.
\subsection{Economic Viability}
\begin{figure}[!t]
\centering
\subfloat[RES-based DERs.]{
  \includegraphics[width=0.9\linewidth]{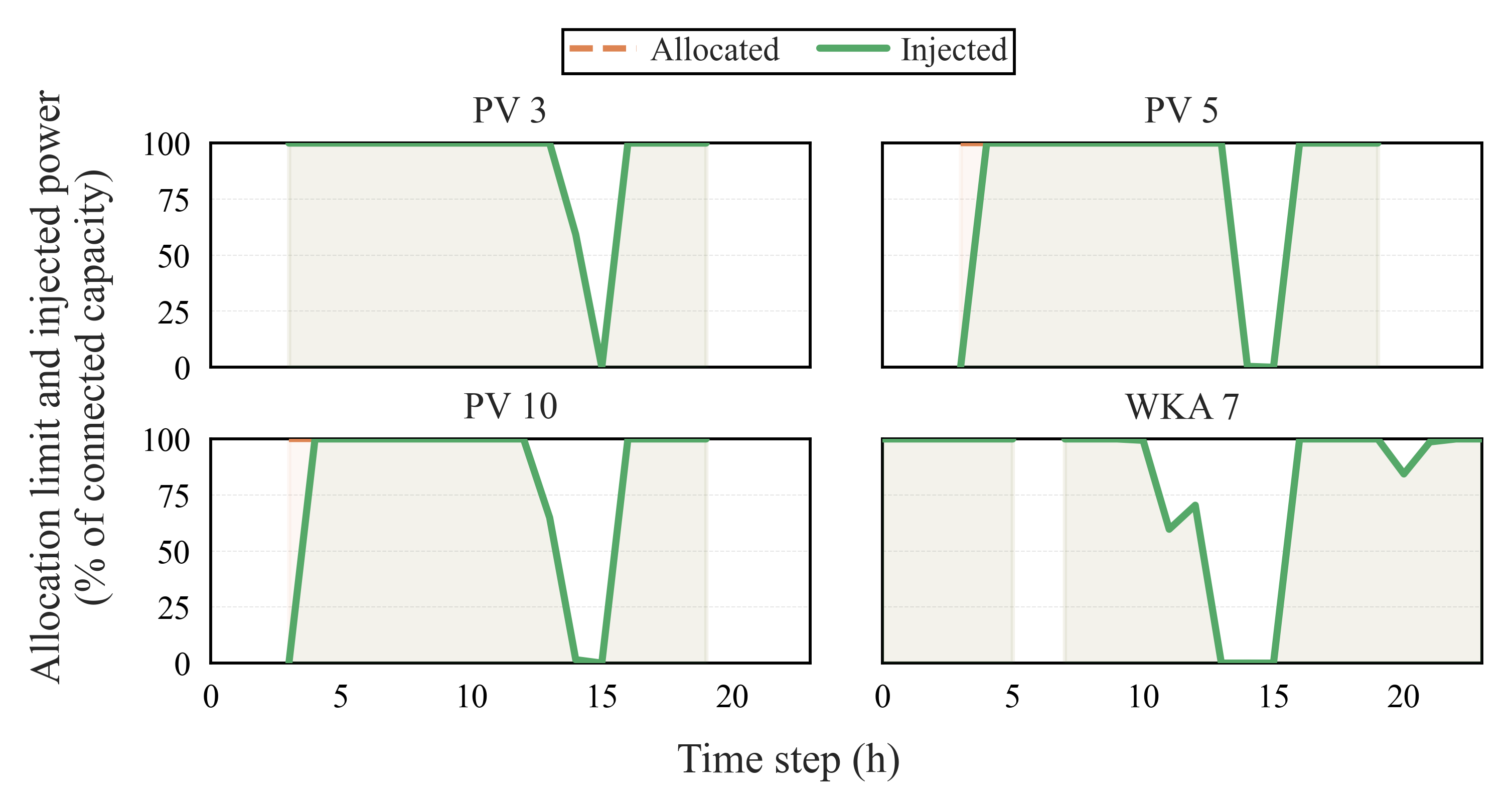}
  \label{fig:allocation_injection_res}
}
\vspace{1mm}

\subfloat[CHP-based DERs.]{
  \includegraphics[width=0.9\linewidth]{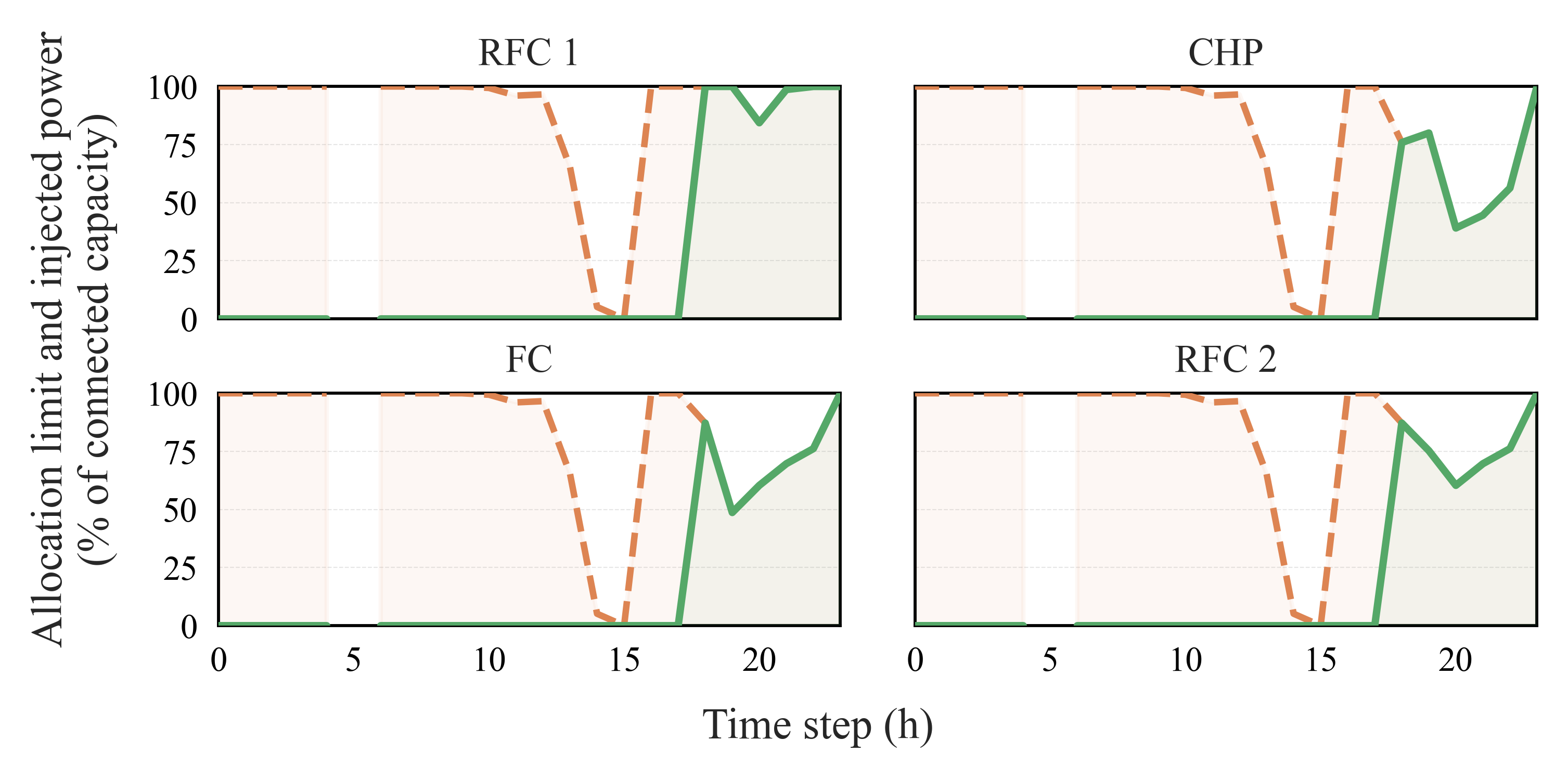}
  \label{fig:allocation_injection_chp}
}
\caption{Temporal profiles of allocated power limits and injected power for representative DERs under the bilevel allocation model, expressed as percentages of connected capacity (time intervals with negligible connected capacity are excluded).}
\label{fig:allocation_injection_combined}
\end{figure}

As shown in Fig.~\ref{fig:allocation_injection_combined}\subref{fig:allocation_injection_res}, RES-based DERs, including PV and wind units, inject power close to their allocated limits, with minor reductions during certain periods caused by grid congestion and the LIFO-based prioritization mechanism. In contrast, Fig.~\ref{fig:allocation_injection_combined}\subref{fig:allocation_injection_chp} shows that CHP-based DERs do not consistently inject power up to their allocated limits. While capacity is allocated based on network feasibility, the actual injection decisions are governed by electricity price dynamics. In particular, CHP injections are reduced or suspended during low-price periods, especially in the early hours of the day, when electricity prices are insufficient to offset fuel costs. Conversely, higher injection levels are observed during peak-price periods in the late afternoon and evening, when market revenues outweigh fuel costs. 
Beyond individual DER behavior, the economic implications are further reflected in the system-level curtailment costs. The bilevel approach achieves the lowest total curtailment cost, amounting to €2{,}946.16, compared to €5{,}951.04 for the OPF-only benchmark and €15{,}049.26 for the LIFO benchmark, corresponding to a reduction of approximately 50\% relative to the OPF-only benchmark and about 80\% relative to LIFO. This improvement is achieved because the bilevel formulation jointly considers network-level allocation decisions and DER-level economic response to electricity prices. As a result, curtailment is allocated more efficiently across both time and DER locations, reducing unnecessary curtailment while maintaining network feasibility. 
\begin{figure}[t]
\centering
\includegraphics[width=0.95\linewidth]{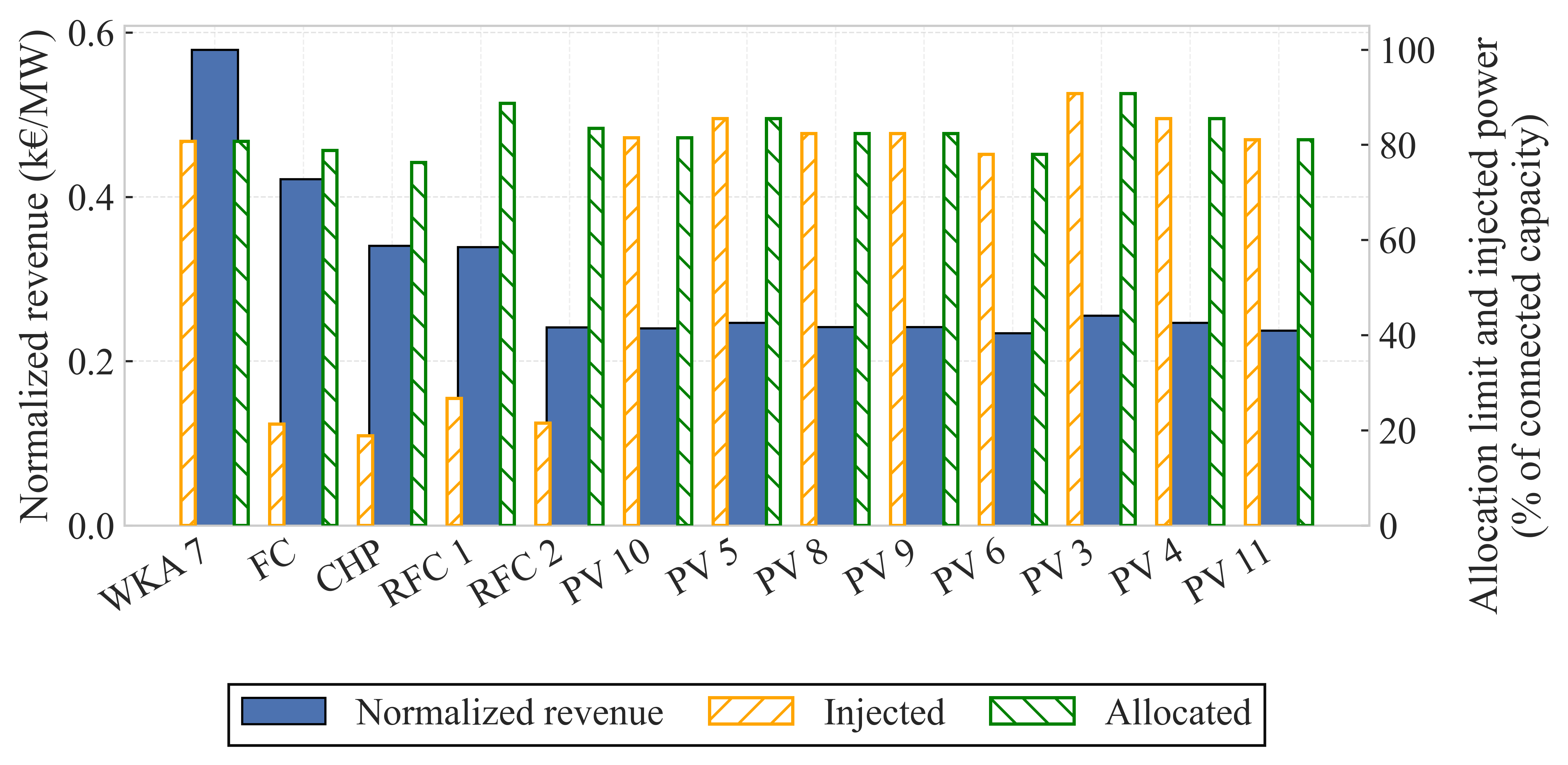}
\caption{Normalized DER revenue and corresponding allocation limit and injected power profiles as a percentage of connected capacity.}
\label{fig:DER revenue and injection percentage}
\vspace{-3mm} 
\end{figure}

\begin{figure}[t]
\centering
\includegraphics[width=0.95\linewidth]{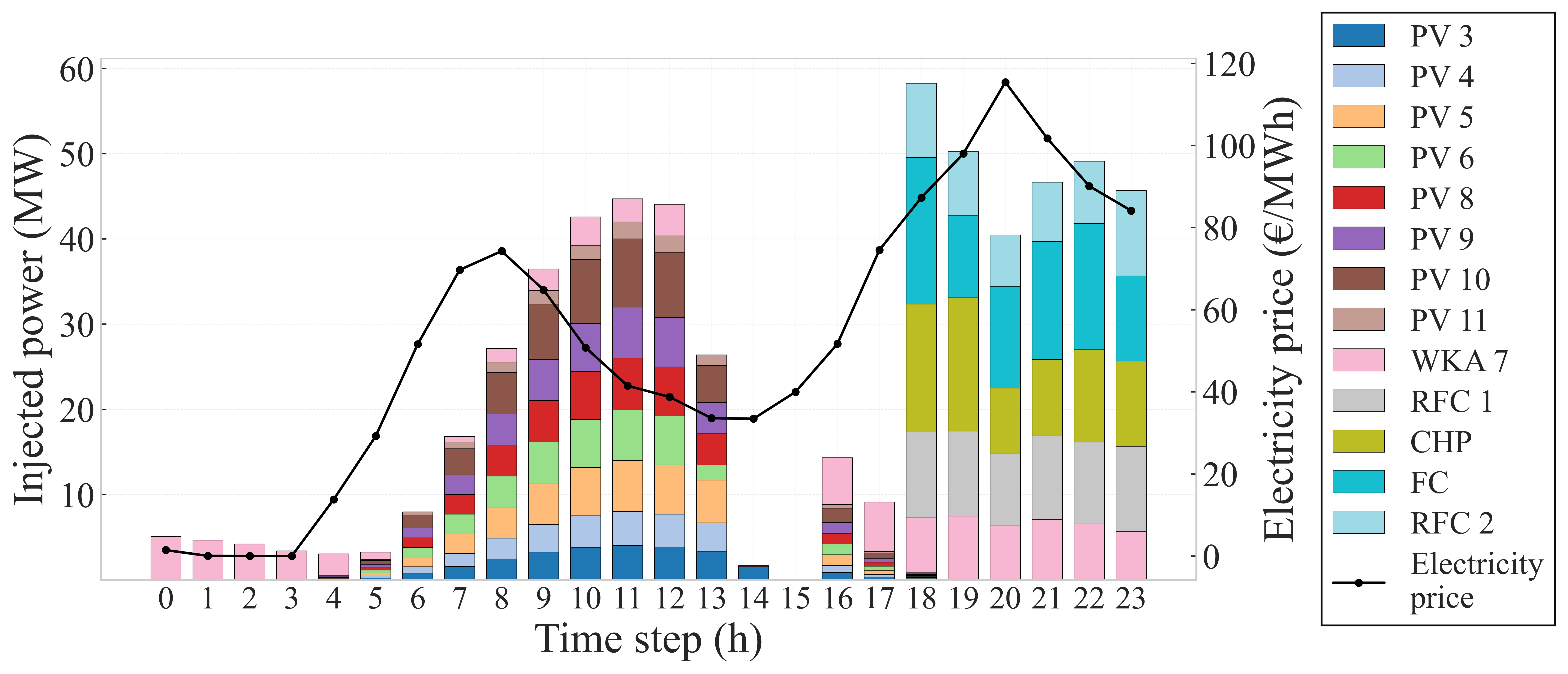}
\caption{Temporal evolution of DER injected power and electricity price.}
\label{fig:DER time_series revenue}
\vspace{-3mm} 
\end{figure}
Figure~\ref{fig:DER revenue and injection percentage} illustrates the normalized revenue of each DER together with the allocation limit and injected power as a percentage of the connected capacity. The normalized revenue is obtained by dividing the DER revenue by its connected (rated) capacity, reported in Table~\ref{tab:load_der_capacities}, enabling a fair comparison among DERs with different sizes and technologies. DERs are ordered from left to right according to decreasing normalized revenue. The figure shows that the normalized allocation limit remains relatively high and consistent across all DERs, indicating that the proposed allocation model provides comparatively balanced access to network capacity. In contrast, the normalized injected power exhibits larger variations due to the different operational characteristics and market participation behaviors of the DERs.

Fuel-based DERs, such as fuel cell and CHP units, exhibit lower normalized injected power than several PV units while still achieving relatively high normalized revenues. This behavior is further illustrated in Fig.~\ref{fig:DER time_series revenue}, which depicts the temporal evolution of DER injected power together with the electricity price profile over the 24-h scheduling horizon. It can be observed that fuel-based DERs primarily inject power during high-price periods, particularly during the evening price peak, enabling them to achieve higher revenues despite lower overall injected power, as also shown in Fig.~\ref{fig:DER revenue and injection percentage}. In contrast, PV units inject larger amounts of power during daytime periods characterized by comparatively lower electricity prices, resulting in lower normalized revenues despite their higher injection levels.
\section{Conclusion}
This paper proposed a dynamic capacity allocation model for distributed energy resources (DERs) operating under non-firm connection agreements. The model explicitly captures the hierarchical interaction between the distribution system operator (DSO), which determines time-varying allocation limits subject to OPF constraints and a Last-In-First-Out (LIFO) prioritization rule, and DER owners, which optimize their power injections in response to economic signals. By embedding a dynamic grid usage penalty signal within the bilevel structure, the model discourages operating conditions close to network technical limits while preserving economically rational dispatch decisions.

The proposed model achieves a balanced and efficient utilization of network capacity while maintaining grid-feasible operation and line loadings within secure operating margins. Compared to the benchmark models, the bilevel model simultaneously reduces the number of curtailment intervals, curtailment severity, and total curtailed capacity. While the LIFO-only model leads to persistent curtailment of lower-priority DERs and the OPF-only model results in irregular and less predictable allocation behavior, the bilevel model achieves a more balanced distribution of curtailment across DERs. The results further show that the bilevel model improves the temporal stability of allocation profiles by mitigating abrupt allocation variations and providing more consistent allocation dynamics across DERs.

From an economic perspective, the proposed bilevel model achieves the lowest overall curtailment cost among all evaluated models while maintaining economically viable revenues for individual DERs. By jointly accounting for network operating constraints, the LIFO-based prioritization rule, and DER responses to electricity price signals, the bilevel model allocates connection capacity more effectively across both time and network locations. This leads to economically efficient DER operation and mitigates excessive curtailment of lower-priority DERs. Consequently, the proposed model supports the practical implementation of non-firm connection agreements by enabling higher DER integration while preserving grid-feasible operation.

\appendices
\ifCLASSOPTIONcaptionsoff
  \newpage
\fi

\end{document}